\documentclass{aa}
\usepackage{amsmath}
\usepackage{graphicx}
\usepackage[utf8]{inputenc}
\usepackage[usenames]{color}
\usepackage{txfonts}
\graphicspath{{Graficos/Hildas/}}

\begin{document}

   \title{Collisional study of Hilda and quasi-Hilda asteroids}
   \author{P. S. Zain\inst{1,2}\thanks{pzain@fcaglp.unlp.edu.ar}, 
          R. P. Di Sisto\inst{1,2} \and R. Gil-Hutton\inst{3}}

   \offprints{Patricio Zain}
  \institute{Instituto de Astrofísica de La Plata, CCT La Plata-CONICET-UNLP. Paseo del Bosque S/N (1900), La Plata, Argentina.                             
   \and Facultad de Ciencias Astronómicas y Geofísicas, Universidad Nacional de La Plata. Paseo del Bosque S/N (1900), La Plata, Argentina.
   \and Grupo de Ciencias Planetarias, Dpto. de Geofísica y Astronomía, FCEFyN, UNSJ - CONICET, Av. J. I. de la Roza 590 oeste, J5402DCS Rivadavia, San Juan, Argentina.
   }

   \date{Received / Accepted}


\abstract
{The Hilda asteroids are located in the outer main belt (MB) in a stable 3:2 mean-motion resonance (MMR) with Jupiter, while the quasi-Hildas have similar orbits but not directly under the effect of the MMR. Moreover, cometary activity has been detected in quasi-Hildas. }
{In this study, we investigate the collisional evolution of Hilda asteroids and apply it to study the cratering on asteroid 334 Chicago, as well as to determine whether impacts between Hildas and quasi-Hildas can serve as a viable mechanism for inducing cometary activity.}
{Using the \texttt{ACDC} code, we simulated the collisional evolution of Hilda asteroids over a period of 4 Gyr. We considered three initial size-frequency distributions (SFD) and two scaling laws for the collisional outcomes and performed a large set of simulations for each scenario which we used to construct median SFDs of the Hilda population. We also derived impactor SFD on asteroid 334 Chicago and used it to calculate the crater SFD on 334 Chicago. Additionally, we evaluated the subcatastrophic impact timescale between Hilda and quasi-Hilda objects.}
{The observed SFD of Hilda asteroids larger than 3 km is best matched by scenarios assuming that such SFD is mostly primordial, implying minimal collisional activity over time. For smaller sizes, although unconstrained, the SFD steepens significantly due to the catastrophic fragmentation of a small number of multikilometer-sized bodies. We determined that the largest impactor on 334 Chicago measures a few kilometers in size, resulting in a maximum crater size of approximately 30 km. Furthermore, the slope of the crater SFD mirrors that of the initial SFD for subkilometric bodies. While impacts events between Hildas and quasi-Hildas can induce observable activity and although stochastic in nature, the timescale of such events exceeds the dynamical lifetime of quasi-Hildas, making them an unlikely primary mechanism for inducing observable activity.}
{}

\authorrunning{P. S. Zain et al.}
\titlerunning{Collisional study of Hilda asteroids}

\maketitle
\section{Introduction}

The asteroid belt, located between the orbits of Mars and Jupiter, offers a significant opportunity for observing, studying, and analyzing the physical and dynamic mechanisms that have shaped the Solar System from its origin to its current state. The distribution of objects within the asteroid belt, particularly their depletion or accumulation in mean motion resonances (MMR) with Jupiter, underscores the predominant influence of Jupiter's dynamics in shaping the structure of this region. The resonant structure of the asteroid belt has been extensively studied in the past (e.g., \cite{nesvorny1997, nesvorny2002}) primarily to comprehend the effects of MMRs and secular resonances on asteroid orbits, aiming to explain conglomerations or gaps within the asteroid belt. Notably, certain MMRs with Jupiter, such as 3:1, 5:2, 7:3, and 2:1, as well as secular resonances such as $\nu_6$ and $\nu_{16}$, represent highly unstable zones that serve as pathways for asteroids to escape from the Main Belt, particularly towards the Near-Earth Object (NEO) region (e.g., \cite{Bottke2002, Granvik2017}). Conversely, in other MMRs with Jupiter, such as 4:3, 3:2, and 1:1, there exist asteroids whose stability persists over the age of the solar system (e.g., \cite{nesvorny1997, Levison1997}).

In particular, the 3:2 MMR with Jupiter hosts the Hilda asteroids, constituting a numerous population whose critical angle $\Phi = 3 \lambda_{\text{J}}-2 \lambda - \bar{\omega}$ librates around $0^{\circ}$, where $\lambda_\text{J}$ denotes the mean longitude of Jupiter, and $\lambda$ and $\bar{\omega}$ represent the mean longitude and longitude of perihelion of the asteroid, respectively. Thus, when a Hilda asteroid is in conjunction with Jupiter, it is at perihelion, positioned far from Jupiter, which acts as a resonant protection mechanism. The Hilda group is centered at a semimajor axis of $a = 3.97$ au and is predominantly concentrated within a narrow fringe approximately $0.12$ au wide around this value. This central region exhibits significant stability; however, it is surrounded by less stable regions where objects may escape from the resonance over the age of the solar system \citep{Ferraz-Mello1998,nesvorny2002,DiSisto2005}. The primary mechanism for the removal of asteroids from the resonance is through collisions among Hildas, which can alter their orbital elements and critical angles, subsequently leading the asteroids into these unstable regions from which they may escape the resonance altogether \citep{gil2000collisional}.

Hilda asteroids are situated at the outer limit of the main asteroid belt (MB) and exhibit physical properties similar to those of Jupiter Trojans. These predominantly belong to taxonomic classes D, P, and C \citep{dahlgren1997, Demeo2013}, with Hildas being predominantly P-type by mass \cite{Demeo2013}. Additionally, they have low albedos \citep{Grav2012}, rendering Hildas darker than Trojans by 15\%–25\% \citep{Romanishin2018}. Through analysis of SDSS colors, \cite{Gil-Hutton2008} and \cite{Wong2017} found a clear bimodal spectral slope distribution among Hildas, delineating two distinct types of objects: those less-red and red.

Due to their composition, which closely resembles that of Centaurs and comets, Hildas represent an interesting population for studying the physical and dynamic connections among various groups of small bodies within the solar system. Particularly noteworthy are the quasi-Hildas located in the vicinity of the Hilda group. These objects, found in non-resonant orbits, exhibit some with cometary activity \citep{Kresak1979, gil2016comet, correa2024population}, leading to their identification as comets. There are currently 26 known active quasi-Hildas. There are 20 reported by \cite{toth2006quasi}, whom included the pre-capture orbits of D/SL9(1) and (2) and the encounter orbit of fragment D/SL9(D/1993 F2-K) and 39P/Oterma (1942 G1) since prior to its close encounter with Jupiter in 1963 it was a quasi-Hilda. The other objects are: 212P \citep{Cheng2013}, (457175) 2008 GO98 \citep{leonard2017cometary,Garcia-Migani2018}, P/2010 H2 \citep{jewitt2020outbursting}, 2008 QZ44 \citep{Chandler2023} and, recently, 482P/2014 VF40 and 484P/2005 XR132 were reported active.

The quasi-Hildas region is part of the region inhabited by Jupiter Family Comets (JFCs). However, this region also serves as an escape route for Hildas \citep{DiSisto2005}, making it a compelling area for studying physical and dynamical behaviors within the solar system. It is therefore reasonable to expect that the quasi-Hildas region hosts both JFCs, originating from Centaurs and the transneptunian region, and Hildas that have escaped from resonance. Detailed physical and dynamical studies of these objects, such as those conducted by \citep{gil2016comet, Garcia-Migani2018, correa2024population}, can provide valuable insights into their origins. However, there remains a possibility that both Hildas and quasi-Hildas could undergo "activation" due to mutual collisions \citep{jewitt2015active}. Given the ice content of Hildas, such activation could result in cometary activity. The discovery of active asteroids or MB comets as interlopers emphasizes the connection between physical and dynamical processes within small body populations.

The collisional evolution and properties of Hildas, as well as outer main belt asteroids, have been explored by \citep{dahlgren1997, gil2000collisional}. These studies revealed a low level of current collisional activity, suggesting that the characteristics of large Hildas and the Size-Frequency Distribution (SFD) are likely primordial. However, due to the significant increase in observed Hildas in recent years and the growing interest in their relationship with quasi-Hildas, we will revisit the collisional study of this population. Therefore, in this study, we propose to investigate the collisional evolution of Hildas and quasi-Hildas with the objective of determining their SFD, analyzing the distributions of impactors and craters on the largest Hilda asteroids, and estimating the impact frequency with quasi-Hildas to assess their potential for activation.

The paper is structured as follows. In Section 2, we present the used observational data and the construction of the observed size-frequency distributions (SFD). In Section 3, we briefly summarize the main properties and input parameters of the \texttt{ACDC} code and the scenarios we will study. In Section 4, we present our results of the collisional evolution of Hildas, the selection of runs, the final SFD, the impactor and crater distributions on the largest Hilda asteroids, and an estimation of the impact frequency with quasi-Hildas. Finally, the conclusions of this work and the discussions on the results are presented in the last section.

\section{Observational data}

We initially aim to construct cumulative SFDs of Hilda and quasi-Hilda asteroids, which we will use to perform the collisional studies in this work. To achieve this, we retrieved the orbital elements and physical characteristics of all asteroids and comets in the outer main belt from the Jet Propulsion Laboratory (JPL)'s Small-Body database query\footnote{\url{https://ssd.jpl.nasa.gov/tools/sbdb_query.html}} with $3.7\le a \le 4.2$ au. This initial sample comprises 6608 objects. 

\begin{figure*}[htp]
\centering
\includegraphics[width=9cm]{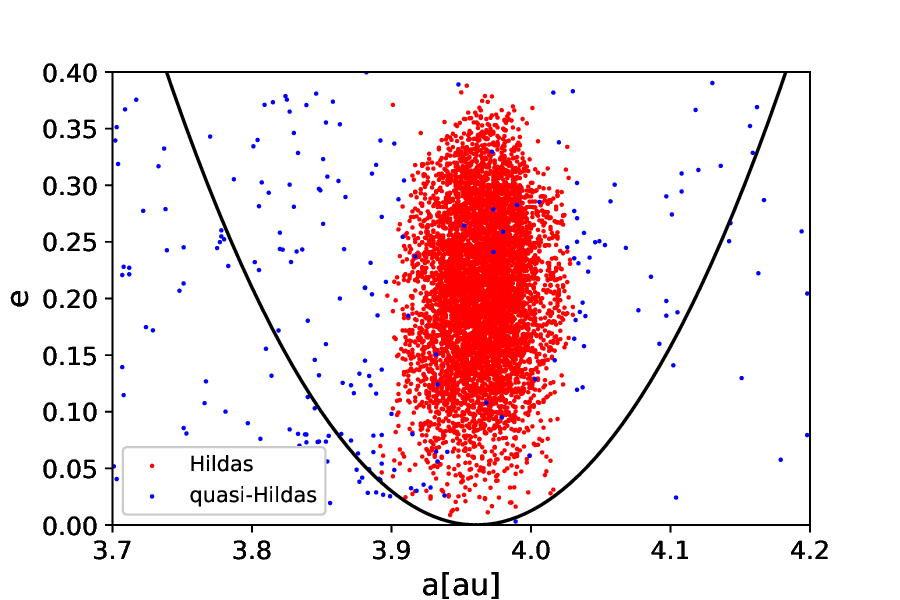}
\includegraphics[width=9cm]{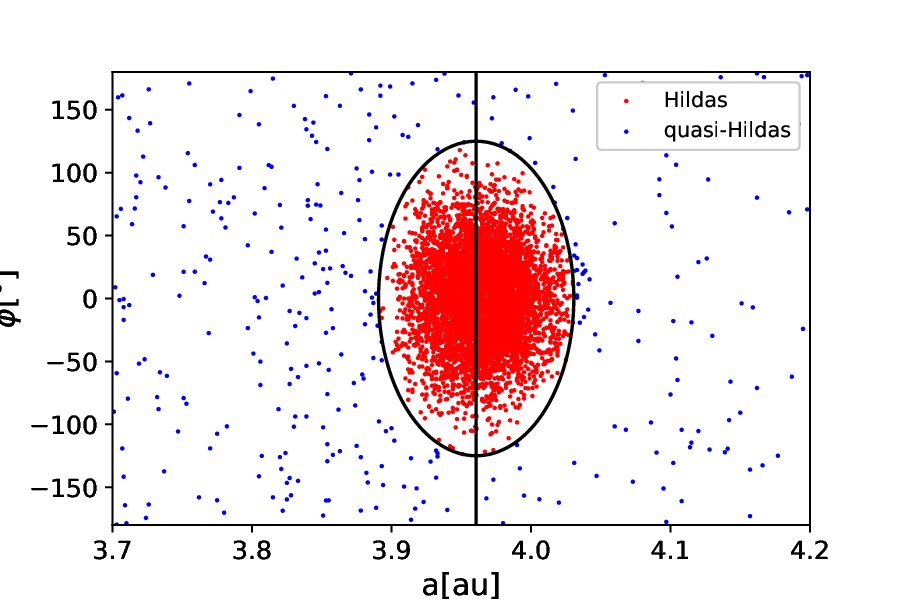}
\caption{Distribution of Hilda asteroids (depicted in red) and quasi-Hilda objects (depicted in blue) presented in the ($a$,$e$) plane (left) and ($a$,$\Phi$) plane. The black ellipse in the left panel indicates the limits between the Hilda and quasi-Hildas regions wheareas the black curve in the right panel indicates the sepparatrix of the 3:2 MMR.}
    \label{fig:aeai}
\end{figure*}

The orbital semimajor axis $a$ and eccentricities $e$ of these bodies are plotted in the left panel of Fig. \ref{fig:aeai}. Almost all of the objects are located near the 3:2 MMR with Jupiter, at $\sim3.9607$ au, and inside the separatrix of the resonance \citep{murray2000solar}, depicted as a black curve. In order to distinguish between Hilda and quasi-Hilda objects, we compute the resonant angle for all bodies:

\begin{equation}
\Phi = 3 \lambda_{\text{J}}-2 \lambda - \bar{\omega},
\end{equation}

where $\lambda_\text{J}$ denotes the mean longitude of Jupiter, and $\lambda$ and $\bar{\omega}$ represent the mean longitude and longitude of perihelion of the asteroid, respectively. The bodies located in the 3:2 MMR have $\Phi$ librating around $0^{\circ}$. The right panel of Fig. \ref{fig:aeai} shows the distribution of $a$ and $\Phi$ for all the bodies. Following criteria similar to \cite{correa2024population}, we define the Hilda population as the bodies located within the ellipse centered at $a=3.9067$ au and $\Phi=0^{\deg}$ with amplitudes of $\Delta a = 0.07$ au and $\Delta \Phi = 125^{\circ}$, a value that corresponds to the $3\sigma$ criteria of the resonant angle distribution. Conversely, we define the quasi-Hildas as the bodies located outside the ellipse in the $(a,\Phi)$ plane.

Although JPL provides diameters and albedos for 1135 objects, the values for the rest are unknown. Thus, we utilize the diameters provided by JPL for bodies larger than 60 km. For smaller bodies and bodies without measured albedo, we calculate the diameters from their absolute magnitude $H$ using the formula \citep{Bowell1989}:

\begin{equation}
D[\text{km]}=\frac{1329}{\sqrt{p_{\textrm{V}}}}10^{-H/5},
\end{equation}

where $p_{\text{v}}$ denotes the visual albedo, which we assume to be 0.055 for Hildas and quasi-Hildas, a value obtained as the median of the albedos of Hilda asteroids provided by JPL. 

The largest asteroid in the region is (334) Chicago, with a diameter of 198.770 km \citep{mainzer2019neowise}, a semi-major axis of approximately $3.899$ au, an eccentricity of approximately $0.025$, and an inclination of approximately $4.68^\circ$\footnote{\url{https://ssd.jpl.nasa.gov/tools/sbdb_lookup.html#/?sstr=chicago}}. (334) Chicago exhibits a significant value of the resonant angle $\Phi$ and frequently encounters Jupiter at close approach distances of approximately 1 au. Moreover, \cite{Ferraz-Mello1998} found that this asteroid's orbit displays chaotic behavior, as the resonant angle alternates between circulation and libration around $0^{\circ}$. It is located in a less stable region of the resonance, which increases the possibility of its escape from the region. However, its low eccentricity makes it highly unlikely to be a captured member of the JFC. Therefore, we consider the asteroid (334) Chicago as a member of the Hilda population, making it the largest body in the region.

\begin{figure}[htp]
\centering
\includegraphics[width=9cm]{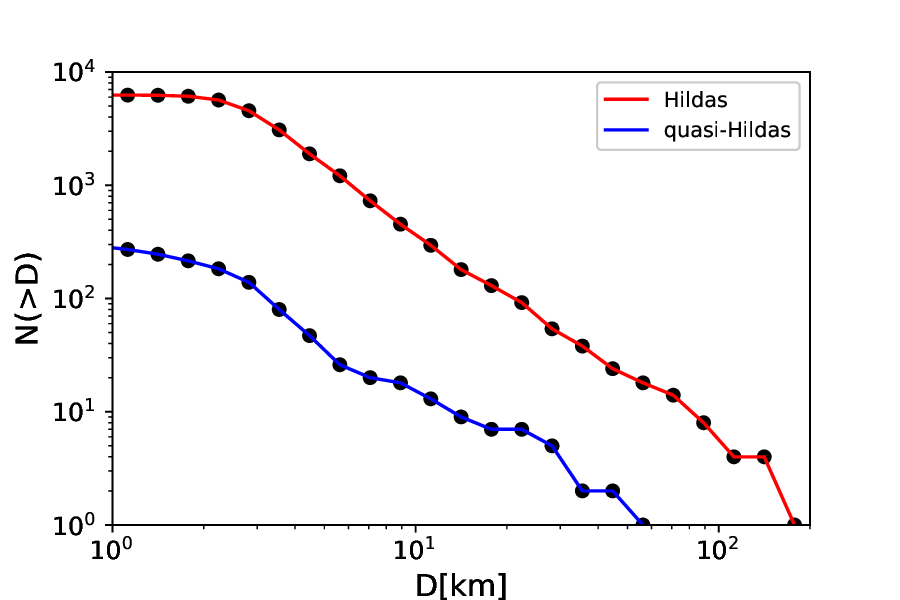}
\caption{Cumulative SFD of the observed Hilda asteroids (red) and the quasi-Hilda objects (blue).}
    \label{fig:ObservedSFD}
\end{figure}

Fig. \ref{fig:ObservedSFD} illustrates the cumulative SFDs obtained for both populations. The SFD of Hilda objects exhibits a cumulative single-slope power law between 3 km and 60 km, with a slope of $-2.03$  consistent with previous findings \citep{terai2018size} and similar to Jupiter's Trojan asteroids \citep{marschall2022implications}. At larger sizes, the slope appears to steepen, although this may be influenced by the limited number of objects. In contrast, the quasi-Hilda SFD does not show a uniform shape due to the reduced sample size, but it can be approximated by a power law with a slope of $-1.35$.


\section{Collisional model}

\subsection{The \texttt{ACDC} code}
\label{sec:ColEVol}

In this study, we employed the \texttt{ACDC} (Asteroid Collisions and Dynamic Computation) code. \texttt{ACDC} is a statistical code designed to simulate the collisional and dynamical evolution of small-body populations over a given integration time, in this case of 4 Gyr. The code was constructed following the outlines of previous collisional evolution models, as \cite{Bottke2005a,Morbidelli2009,OBrien2005,deElia2007,Cibulkova2014}. In this section, we provide a concise overview of the fundamental aspects of the \texttt{ACDC} code. For an in-depth understanding of the construction and implementation of the model, we direct the reader to \cite{Zain2020}. 

The collisional component of the model is determined by tracking changes in the number of bodies resulting from objects being destroyed and fragments being ejected during collisions. In each timestep, \texttt{ACDC} calculates the number of collisions between all pairs of target-projectile bodies based on their intrinsic impact probabilities. Moreover, the occurrence of large impacts follows Poisson statistics. Subsequently, \texttt{ACDC} distributes the fragments created in each event over the different size bins and removes bodies that experience catastrophic disruption.

The dynamical component of the model, as implemented in previous works \citep{Zain2020,zain2021impacts}, incorporates the collective influence of the Yarkovsky effect and resonances. This mechanism serves to remove subkilometric asteroids from a specified region of the MB. However, the population examined in this study resides within a narrow region, approximately 0.07 au in half-width, surrounding the 3:2 MMR with Jupiter. After conducting numerous tests, we found that the inclusion of the Yarkovsky effect leads to rapid depletion of this region which appears to be independent of the physical, dynamical, and geometric properties of the asteroids within the studied area. Conversely, \cite{brovz2008asteroid} investigated the dynamical evolution of asteroids initially positioned within the 3:2 MMR. Their findings suggest that the Yarkovsky effect within a strong first-order MMR primarily influences changes in orbital eccentricity, while the semimajor axis tends to follow the resonance center. Given these considerations, we chose not to incorporate the Yarkovsky effect in this investigation.

We used \texttt{ACDC} to study the collisional evolution of the Hilda population. To do so, we considered collisions between targets and projectiles exclusively within the Hilda region, using the values for the intrinsic collision probability and impact velocity of $1.93 \times 10^{-18}$ yr$^{-1}$ km$^{-2}$ and $3.36$ km s$^{-1}$, respectively \citep{dell2001updated}. 

\subsection{Scaling laws}

The fundamental determinant of collision outcomes is the kinetic energy of the impact. Specifically, we employ the specific impact energy of the projectile:

\begin{equation}
    Q=\frac{1}{2}\frac{m_{j} v_{\text{imp}}^{2}}{\left(m_{i}+m_{j}\right)},
\end{equation}

where $m_{i}$ and $m_{j}$ represent the masses of the target and impactor, respectively, and $v_{\text{imp}}$ is the mutual impact velocity. 

The code compares $Q$ with a function $Q_{\text{D}}^{*}$, defined as the specific energy needed to disrupt and disperse 50\% of the target mass. If a collision results in $Q<Q_{\text{D}}^{*}$, it is labeled as a cratering event, while if $Q \geq Q_{\text{D}}^{*}$, it is classified as a catastrophic disruption event. In either case, the code calculates the masses of the largest remnant, the largest fragment, and the cumulative SFD of the fragments following the relations derived by \cite{BenzAsphaug} and \cite{Durda2007}. The collisional strength of the Hildas, represented by the function $Q_{\text{D}}^{*}$, is an unknown parameter. Therefore, in this study we adopt the functions derived by \cite{BenzAsphaug} for basaltic and icy targets at impact speeds of 5 km s$^{-1}$ and 3 km s$^{-1}$, respectively. 

\subsection{Initial Hilda SFD}

The shape of the initial SFD of the Hilda asteroids is also an unknown property. However, insights from previous collisional and dynamical studies suggest that Hilda populations may be very old and that the collisional activity has been very low \citep{gil2000collisional,davis2002collisional,brovz2011did}. Consequently, we assume that the SFD for sizes larger than 10 km is primordial. 

For smaller sizes, we adopt the approach outlined in \cite{marschall2022implications} in their collisional study of Jupiter's  Trojan asteroids,  due to their physical similarities and common origin with Hilda asteroids \citep{terai2018size}. We propose three distinct scenarios for the initial Hilda SFD: S32, where the initial slope below 3 km continues the observed slope between 3-10 km; S31, where the slope below 3 km is set to 1; and S10, where the slope below 10 km is also set to 1, resulting in a shallower slope than the observed population. For each scenario, we conducted two sets of 20,000 simulations, employing the scaling laws for icy and basaltic targets as described by \cite{BenzAsphaug}.

\begin{figure}
    \centering
    \includegraphics[width=8cm]{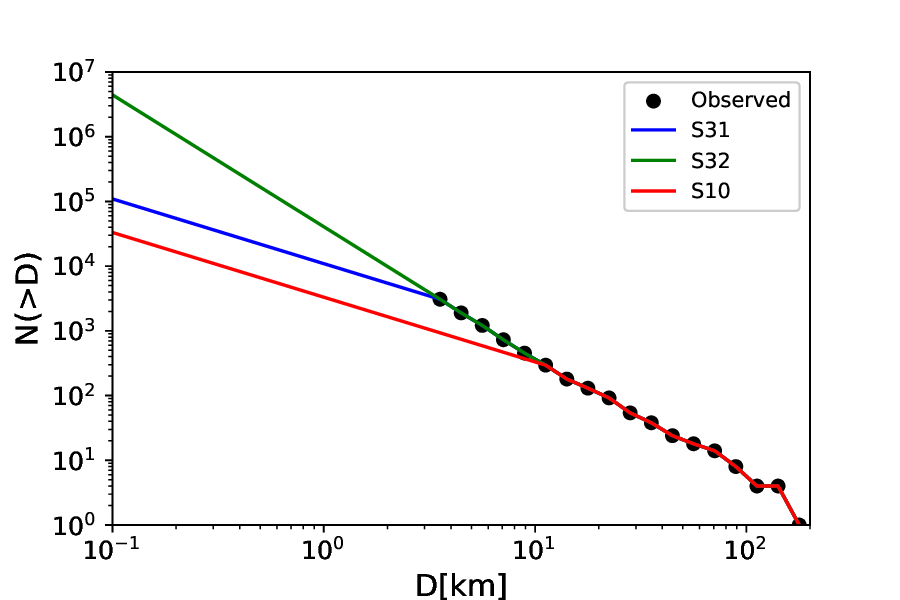}
    \caption{Initial SFD tested for the Hilda asteroids, labeled as S31, S10 and S32. Black dots indicate the observed SFD for Hilda asteroids larger than 3 km. }
    \label{fig:InitialSFD}
\end{figure}

\section{Results}

\subsection{Collisional evolution of Hildas}

\begin{figure}[h]
    \centering
\includegraphics[width=8cm]{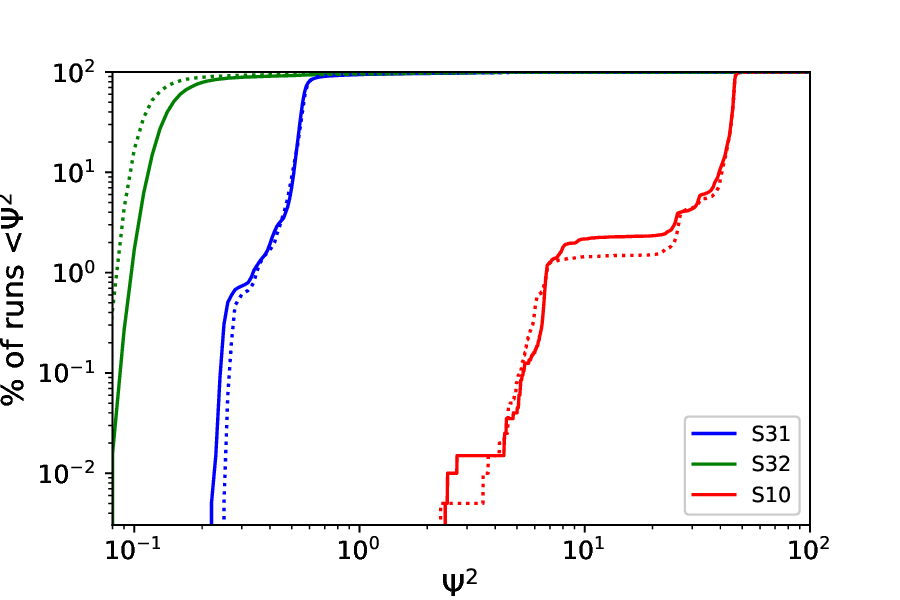}
    \caption{Distribution of the $\psi^{2}$ across 20,000 runs in each scenario. The x-axis represents $\psi^{2}$ values, while the y-axis depicts the percentage of runs with values below the corresponding $\psi^{2}$ threshold. Solid lines indicate the results for the Ice scaling law, while dashed lines represent the results for the Basalt scaling law.}
    \label{fig:Metric}
\end{figure}

In this section we present the results on the collisional evolution of the Hilda population through 4 Gyr. One of the main features of \texttt{ACDC} is its high level of stochasticity, treating large impacts as random events through the application of Poisson statistics. As a result, runs using different random seeds may produce varied results. To address this variability, we perform an extensive set of 20000 runs for each scenario and for each scaling law. Then we apply observational constraints to identify and interpret the most robust results statistically. In this particular study, the only observational constraint considered is the Hilda SFD for sizes larger than 3 km, as shown in Fig. \ref{fig:ObservedSFD}. 

To quantitatively assess the fidelity of a simulation in replicating observational data, we follow the methodology outlined by \cite{Bottke2005a}. We define a metric designed to evaluate the goodness of fit between the observed SFD, denoted as $N_{\text{obs}}$, and the simulated SFD, denoted as $N_{\text{sim}}$:

\begin{equation}
\psi^{2} = \sum_{i}\left(\frac{N_{\text{sim}}\left(>D_{i}\right) - N_{\text{obs}}\left(>D_{i}\right)}{0.2 N_{\text{obs}}\left(>D_{i}\right)}\right)^{2},
\end{equation}

where the summation extends over size bins ranging from 3 km to 100 km. We calculate $\psi^2$ for each simulation, where a lower value of $\psi^2$ indicates better alignment with the observational data. Fig. \ref{fig:Metric} illustrates the distribution of metrics among the runs performed in each scenario. In particular, scenario S32 exhibits the lowest metrics, with 80\% of runs yielding $\psi^2$ values below 0.2. Following closely is scenario S31, where 80\% of runs result in $\psi^2$ values smaller than 1. For comparison, \cite{Bottke2005a} suggests that a simulation provides a good fit if $\psi^2<20$. The S10 scenario is capable of producing good fits but in a much smaller quantity. We find that approximately 1\% of the runs produce good fits according to the criteria of \cite{Bottke2005a}, with values of $\psi^2$ between 2 and 20. In all cases, we observe that the distribution of the metrics remains consistent regardless of whether we consider $Q_\text{D}^{*}$ for icy or basaltic targets.

We utilize the $\psi^2$ metrics to select the runs that offer the best fits. The SFD obtained from these runs is used to construct a representative median SFD for sizes ranging from 10 cm to 200 km, by calculating the median number of objects in each size bin using the individual SFDs that meet the given selection criteria. For scenarios S31 and S32, we include simulations with $\psi^2\le1$, while for the S10 scenario we include those with $\psi^2<20$, following the criteria of \cite{Bottke2005a}.

\begin{figure}
    \centering
    \includegraphics[width=8cm]{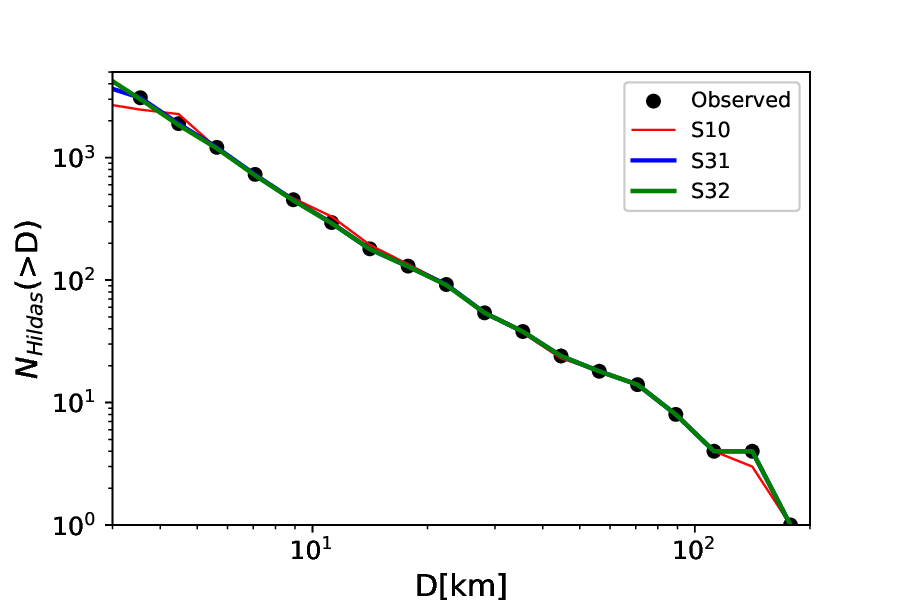}
    \caption{Median SFD of Hilda asteroids larger than 1 km after 4 Gyr of collisional evolution for the S10 (red), S31 (blue), and S32 (green) scenarios. Black dots indicate the observed SFD km. }
    \label{fig:FinalSFD}
\end{figure}

The median SFDs are depicted in Fig. \ref{fig:FinalSFD}. We note that scenarios S31 and S32 exhibit nearly exact fits with the observed SFD for sizes larger than 3 km. These scenarios were formulated under the assumption that the observed SFD in that size range is primordial, suggesting that minimal catastrophic collisional activity occurred in the multikilometer-sized Hilda asteroids over 4 Gyr. The S10 scenario, which didn't initially match the observed SFD, was still able to generate an imperfect but acceptable fit within the observed range. However, it is worth recalling that this median was constructed using significantly fewer runs than the other two scenarios. 

\begin{figure*}
    \centering
   \includegraphics[width=8cm]{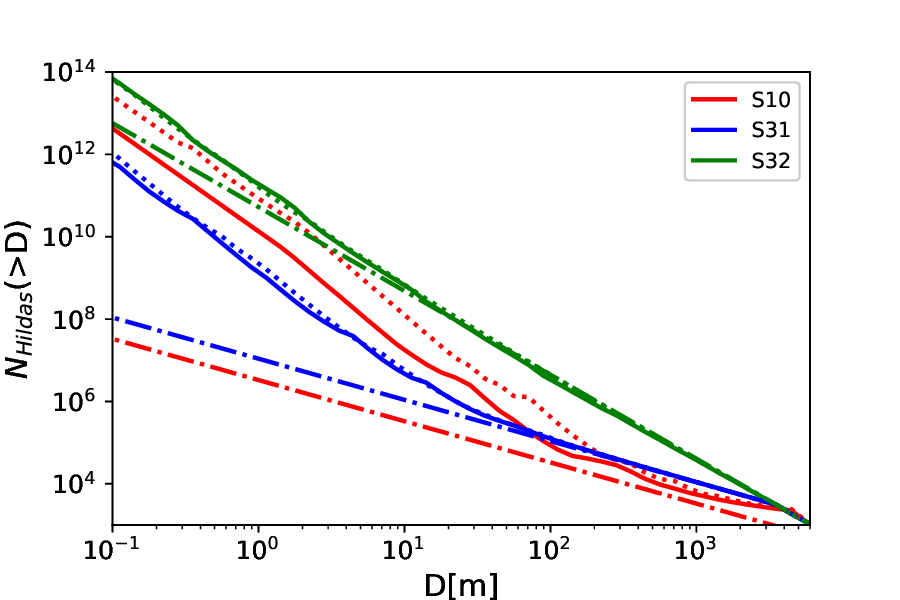}
   \includegraphics[width=8cm]{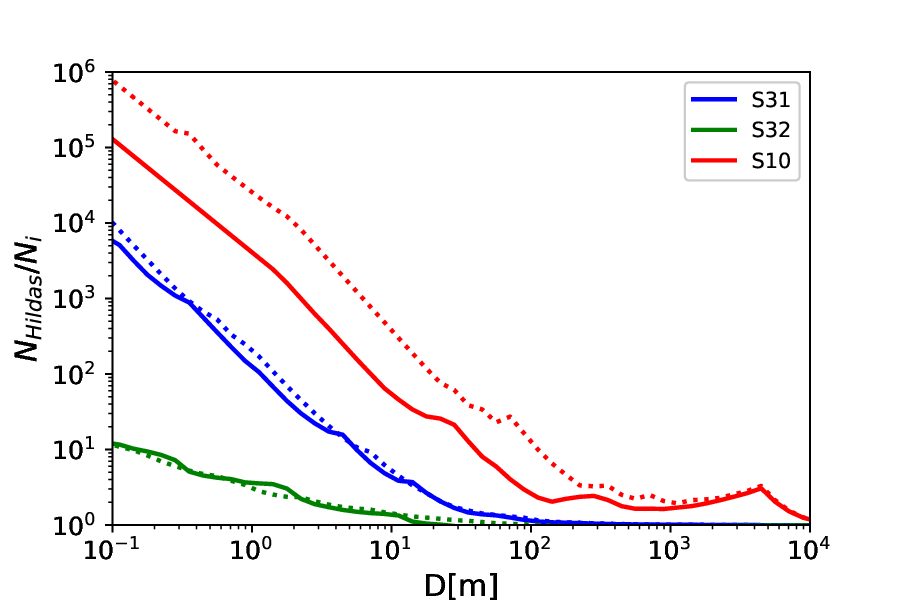}
   \includegraphics[width=8cm]{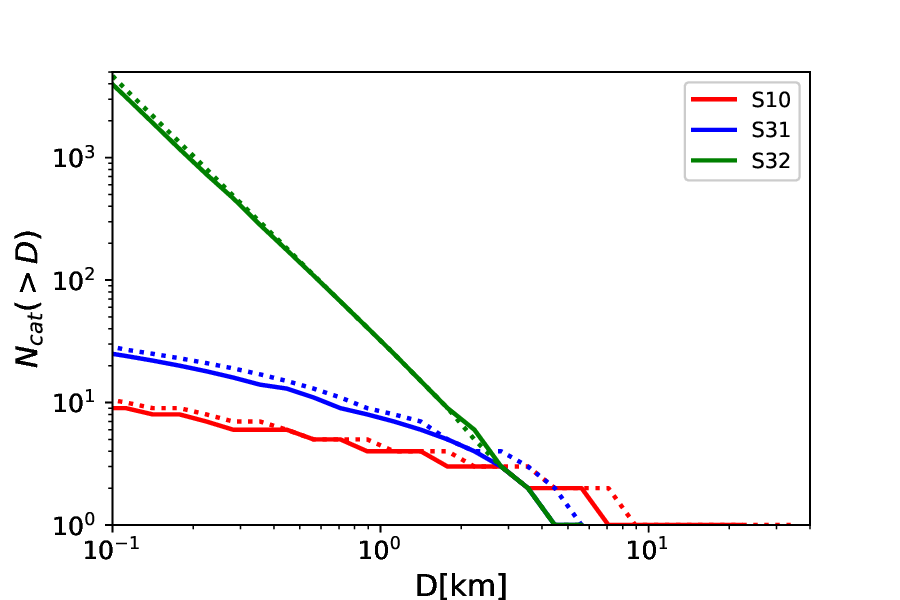}
   \includegraphics[width=8cm]{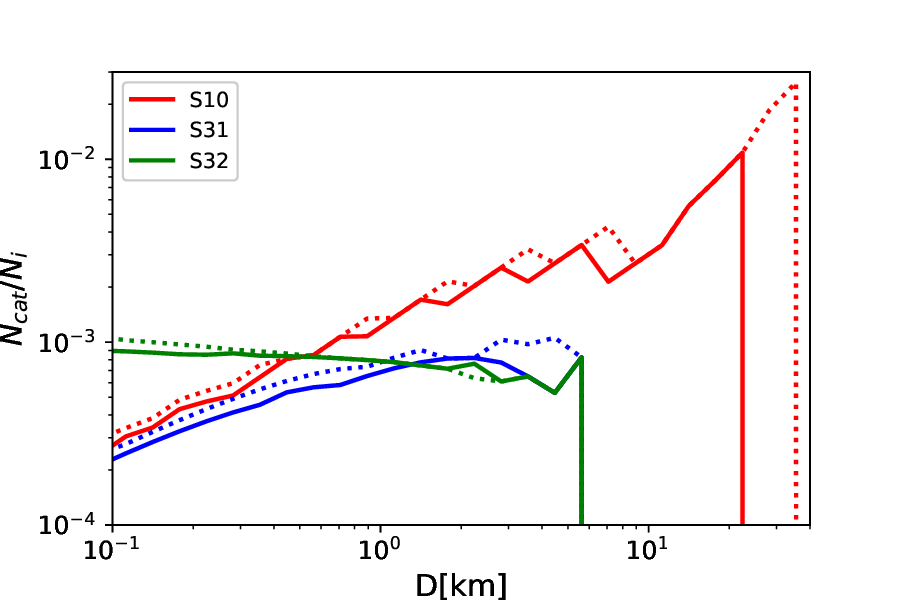}
        
    \caption{Top: Median SFD of Hilda asteroids (left) and ratio with respect to the initial SFD in the subkilometer size range for the S10 (red), S31 (blue), and S32 (green) scenarios. Dashdotted lines represent the initial SFD. 
    \\ Bottom: SFD of catastrophically disrupted bodies (left) and ratio of catastrophically disrupted bodies with respect to the initial SFD (right) for the S10 (red), S31 (blue) and S32 (green) scenarios. 
    \\ Solid lines represent simulations conducted with icy targets scaling laws, while dashed lines indicate simulations with basaltic targets.}
        \label{fig:ChicosSFD}
\end{figure*}

The median SFD diverges from the initial populations in the subkilometer size range, where we lack observational constraints to distinguish between one and another. However, we can utilize our median SFDs in the mentioned size range to address the general behavior. The top left panel of Fig. \ref{fig:ChicosSFD} illustrates the median SFD with icy and basaltic targets. Remarkably, the distributions obtained with the different scaling laws are nearly identical for the S31 and S32 scenarios, possibly due to the generally low collisional activity in the region. In the S31 scenario, both SFD distributions exhibit slight differences in the meter-sized range, where the SFD for icy bodies is slightly lower than the SFD for basaltic bodies. Furthermore, all scenarios exhibit steeper slopes than the initial populations, implying a general increase in the number of small bodies. These slopes have values of 2.59 and 2.54 for S31 and 32, respectively, indicating a state of near collisional equilibrium \citep{dohnanyi1969}. This increase in the SFD is further highlighted in the top right panel of Fig. \ref{fig:ChicosSFD}, which presents the ratio between the median and the initial SFD. In the S32 scenario, the median SFD is approximately 10 times higher than the initial SFD for bodies larger than 10 cm and decreases to 1 for bodies larger than 20 m for icy targets and 100 m for basaltic targets. The increase in numbers is much more pronounced in the S31 scenario, where the ratio increases to $10^4$ for bodies larger than 10 cm and decreases to 1 in the kilometer range. Conversely, Fig. \ref{fig:ChicosSFD} illustrates that the collisional activity was more intense in the S10 scenario. On one hand, the left panel reveals that the median SFD exhibits a more pronounced wavy shape compared to the other scenarios. On the other hand, the right panel shows that the ratio between the median and initial SFD extends to multikilometer-sized bodies, and increases with decreasing size until reaching $10^6$ for cm-sized bodies. Furthermore, the SFDs and ratios obtained with the different scaling laws exhibit differences of over an order of magnitude.

The model utilized in this study lacks any dynamical mechanism for the removal of small bodies. Consequently, the increase in the number of bodies suggests that the Hilda SFD in the subkilometer range is solely influenced by the catastrophic fragmentation of larger bodies. The bottom left panel of Fig. \ref{fig:ChicosSFD} illustrates the SFD of catastrophically disrupted bodies larger than 100 m, while the bottom right panel shows the ratio of catastrophically disrupted bodies with respect to the initial SFD. We observe that the highest number of catastrophic collisions occur in the S32 scenario, resulting in $3\times10^{3}$ destroyed bodies larger than 100 m, which represent only $10^{-3}$ of the initial SFD. Conversely, the number of catastrophic collisions is much lower in the S31 scenario, where only 25 and 10 bodies larger than 100 m and 1 km, respectively, are disrupted, representing up to $\sim 10^{-3}$ of the initial SFD. In both the S31 and S32 scenarios, the largest catastrophically disrupted body has a size of 6 km, and the results are very similar regardless of the scaling law considered. The number of catastrophic collisions is even smaller for the S10 scenario, where only 10 bodies larger than 100 m are catastrophically disrupted. However, since this scenario is much less numerous in general terms, the catastrophic collisions represent a larger fraction of the initial population, reaching values up to $3\times10^{-2}$. 
The number of catastrophic collisions is even smaller for the S10 scenario, where only 10 bodies larger than 100 m are catastrophically disrupted. Additionally, we find that the S10 scenario results in the disruption of larger bodies than the other two scenarios, with the largest disrupted body having a size of 22 km and 35 km for the ice and basalt scaling laws, respectively. Indeed, as the number of catastrophic collisions is low, this results in a higher number of projectiles in the SFD capable of disrupting larger bodies.


\subsection{Impactors and cratering on the largest Hilda asteroid}

Using \texttt{ACDC}, it is possible to record the number of collisions of projectiles on bodies from a given size bin during the integration time of 4 Gyr. We utilize this to obtain a median SFD of impactors larger than 1 m on the largest asteroid in the Hilda region, using the runs selected in the previous section. This particular asteroid was discussed in Section 2 and is called 334 Chicago. It has a diameter of approximately $\sim 200$ km and is classified as a C-type according to Tholen  taxonomy. We obtain that this asteroid was not hit by large bodies. In particular, the largest impactor has a size of 1.77, 1.41 km and 2.24 km for S10, S31 and S32 scenarios, respectively.

\begin{figure}
    \centering
    \includegraphics[width=8cm]{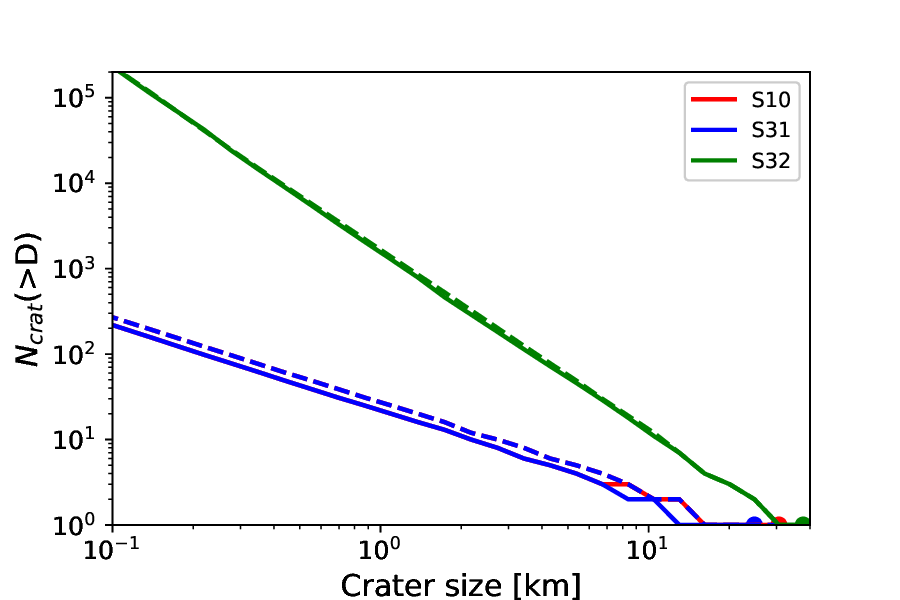}
    \caption{SFD of craters on the surface of asteroid 334 Chicago, the largest member of the Hilda asteroids, for the S10 (red), S31 (blue) and S32 (green) scenarios. Circles indicate the size of the largest crater. Solid lines depict the results for the Ice scaling law, while dashed lines represent the results for the Basalt scaling law.}    
    \label{fig:ImpactorSFD}
\end{figure}

With this information, we can derive an SFD for the number of craters produced by these impacts. To calculate the apparent transient crater diameter $d_t$ resulting from a projectile of diameter $d$, we employ the scaling law proposed by \cite{Holsapple2007}:

\begin{equation}
d_{\text{t}}=K_{1}\left[ \left( \frac{gd}{2v_{i}^{2}}\right)\left(\frac{\rho_{t}}{\rho_{i}}\right)^{\frac{2\nu}{\mu}}+K_{2}\left(\frac{Y}{\rho_{t}v_{i}^{2}} \right)^{\frac{2+\mu}{\mu}} \left(\frac{\rho_{t}}{\rho_{i}}\right)^{\frac{\nu \left(2+\mu\right)}{\mu}}  \right]^{-\frac{\mu}{2+\mu}}d
\label{eq:Crater}
\end{equation}
where $\rho_{\text{t}}=1.5$ g cm$^{-3}$ is the target density, $g$ its surface gravity, $Y$ its strength, $\rho_{\text{i}}$ the density of the impactor which we assume the same as $\rho_\text{t}$, and $v_{i}$ the impactor velocity. The two exponents $\mu$ and $\nu$ and the constants $K_1$  and $K_2$ characterize the target material. Following our previous work where we calculated the crater SFD for Ceres \citep{zain2021impacts}, we use $K_1=1.67$, $K_2=0.8$, $\mu=0.38$ and $\nu=0.378$, and $Y=4\times 10^6$ dyn cm$^{-2}$\citep{Kraus2011,DiSisto2013,Hiesinger2016}. The final crater size is obtained by multiplying eq. \ref{eq:Crater} by $1.3 k$, with $k=1.19$ \citep{Marchi2011}. 

The crater SFDs for the three scenarios are shown in Fig. \ref{fig:ImpactorSFD}. The largest craters, represented as dots in the figure, measure 30 km, 24 km, and 37 km in the S10, S31, and S32 scenarios, respectively. Remarkably, the crater distributions exhibit notable differences. While the S10 and S31 distributions overlap across all size ranges, the S32 scenario displays a considerably higher number of craters, spanning several orders of magnitude compared to the other two scenarios. However, it is noteworthy that the slope observed in the three crater distributions mirrors the slope of their respective initial SFDs for subkilometer bodies. This observation could provide a valuable constraint for future observational efforts aimed at determining the initial distribution of Hilda asteroids, should a space mission targeting Hilda asteroids be developed. Nevertheless, it is essential to emphasize that this study does not account for possible impacts with bodies from other populations, such as the Main Belt, Jupiter Trojans, Jupiter Family Comets, or early planetesimals. Also, we neglect the effect of crater saturation, as the small number of impacts makes it unlikely for the surfaces to become saturated.

\subsection{Impacts and observable activity on quasi-Hildas}

In this section, we discuss the contribution of impacts to quasi-Hildas in terms of their observable activity. To accomplish this, we utilize the expression derived by \cite{jewitt2015active}, which provides the minimum radius of a projectile $r_\text{p}$ required to create ejecta with a substantial cross section when impacting a target of radius $r_\text{t}$.

\begin{equation}
    r_\text{p}=\left(\frac{3af}{4A}\right)^{1/3}v_{\text{esc}}^{-\alpha/6} U^{\alpha/3} r_{\text{t}}^{(2-\alpha)/3},
    \label{eq:rmin}
\end{equation}

where $v_{\text{esc}}$ denotes the escape velocity of the asteroid, $U=3.34$ km s$^{-1}$ represents the impact velocity \citep{dell2001updated}, $A=0.01$ and $\alpha=-1.5$ are parameters from \citep{housen2011ejecta}, and $a=0.1$ mm is specified by \cite{jewitt2015active}. The parameter $f$ relates the mass of the ejecta to the scattering cross-section and also to changes in the apparent magnitude. We adopt a value of $f=1$, which, as indicated by \cite{jewitt2015active}, would lead to an approximate doubling of the asteroid's brightness, making it readily detectable. According to \cite{jewitt2015active}, even modest impacts can produce observable ejecta signatures. In fact, employing Eq. \ref{eq:rmin}, we find that an impact of a 30 cm projectile onto a 1 km asteroid would eject enough material to double the brightness post-impact. Similarly, a 45 m projectile would have the same effect by impacting a 10 km asteroid.

We estimate the occurrence of impact-triggered quasi-Hilda activity by calculating the frequency of subcatastrophic collisions between quasi-Hildas and Hildas. To do so, we make the simplifying assumption that both populations share the same impact speeds and probabilities. We consider targets from the quasi-Hilda population, with sizes ranging from 1 km to 14 km. The projectiles are Hilda asteroids, which follow the SFD obtained from collisional evolution in this study in the different scenarios. We utilize the SFDs obtained using the ice scaling law, as Eq. \ref{eq:rmin} requires that both the target and the projectile have the same density. The frequency is then calculated as:

\begin{equation}
    frec_\text{t} = \frac{P}{4} \sum_\text{p} \left(D_\text{t}+D_\text{p}\right)^2 N_\text{p},
    \label{eq:frecuencias}
\end{equation}

where $P=1.93 \times 10^{-18}$ yr$^{-1}$ km$^{-2}$ is the intrinsic collision probability \citep{dell2001updated}, and $N_\text{p}$ denotes the number of Hilda projectiles with diameter $D_\text{p}$. The summation extends from a minimum value of $2r_p$ (as per Eq. \ref{eq:rmin}) to a maximum size of $(2Q^{*}/U^2)^{1/3}D_t$, which corresponds to the projectile size capable of catastrophically disrupting the targets \citep{Bottke2005a}.

\begin{figure}
    \centering
    \includegraphics[width=8cm]{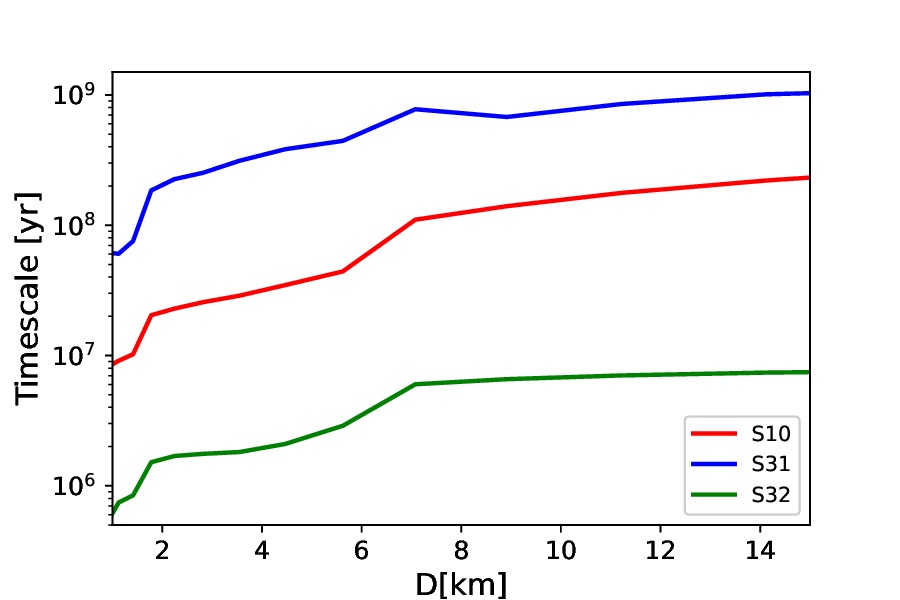}
    \includegraphics[width=8cm]{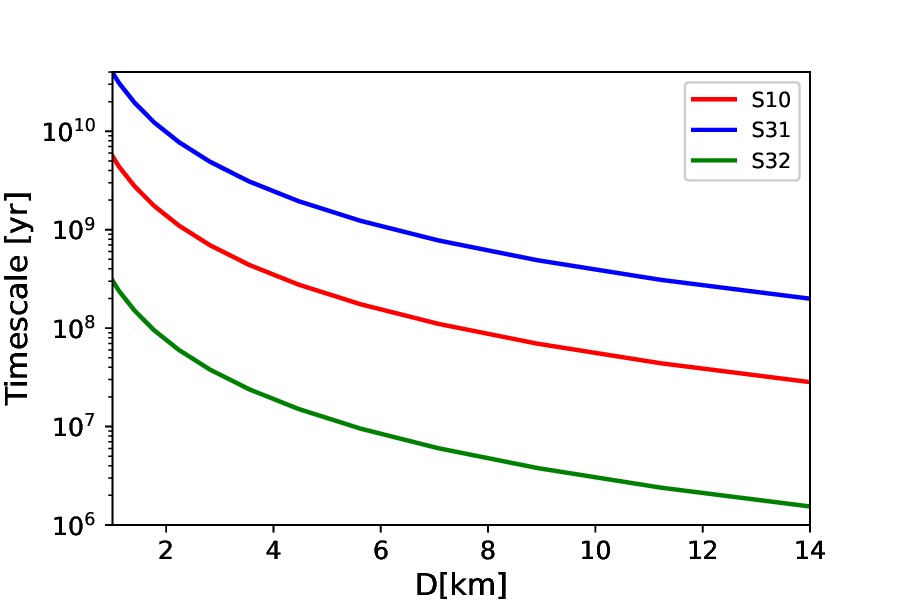}
    \caption{Timescale of subcatastrophic impacts capable of producing observable activity on quasi-Hildas for the S10 (red), S31 (green) and S32 (blue) scenarios. The top panel indicates the timescale for impacts onto a single quasi-Hilda target. while the bottom panel indicates the timescale of impacts from projectiles capable of forming a 10 m deep crater and exposing the ice beneath the surface.}
    \label{fig:FrecQH}
\end{figure}

We plot in the top panel of Fig. \ref{fig:FrecQH} the impact timescales calculated as the inverse of the frequencies calculated from eq. \ref{eq:frecuencias}. It is evident that the difference in the timescales between each scenario is approximately an order of magnitude, reflecting the difference in the number of bodies at the smaller end of the SFDs. The timescales are shorter in the S32 scenario, where the SFD in Fig. \ref{fig:ChicosSFD} shows almost $10^{14}$ objects larger than 10 cm. Consequently, a quasi-Hilda object with size between 2 km and 4 km experience one impact every $1.6 \times 10^{6}$ years, increasing to $5 \times 10^{6}$ years for targets larger than 7 km. The impact timescales exhibit a similar trend, with one and two orders of magnitude larger timescales for the S10 and S31 scenarios, respectively.


It is worth noting that while the frequencies and timescales are calculated based on deterministic laws, the occurrence of impacts on quasi-Hilda asteroids is stochastic due to the low number of objects in these regions. Specifically, we only have 20 objects larger than 10 km in the quasi-Hilda region. Nevertheless, we find that these timescales are many orders of magnitude longer than the dynamical lifetimes of objects in the quasi-Hilda region. We calculated from previous studies that objects, whether escaped Hildas as described in \cite{DiSisto2005} or JFCs as mentioned in \cite{Disisto2009}, spend a few thousand years traversing the region, eventually reaching up to approximately $5 \times 10^5$ years. Additionally, \cite{Garcia-Migani2018} identified the object (457175) as a quasi-Hilda for a duration of $7\times10^4$ years. Overall, our analysis indicates that the frequency of impacts capable of inducing activity on quasi-Hildas is too low for collisions between Hildas and quasi-Hildas to be deemed a primary mechanism. 

Nevertheless, it should be noted that the vast majority of quasi-Hildas observed to be active display their activity in the post-perihelion arc. This suggests that thermal processes primarily trigger activity and that it generally does not initiate solely due to the thermal inertia of the object. In other words, the collision of a projectile causing fragments to increase the reflective surface and thus brightening may not be the preferential process \citep{Vincent2015}. Cometary ice transforms into compact amorphous ice down to depths of approximately 10–20 m due to galactic cosmic rays \citep{Maggiolo2020}. Cosmic rays also contribute to the formation of a refractory mantle, requiring the outermost layer of a comet to be broken to expose ices and facilitate sublimation. Consequently, collisions occurring in quasi-Hildas could potentially disrupt or penetrate the outermost layer, leaving the comet's ice vulnerable. Subsequently, upon passing through perihelion, the comet resumes its thermal sublimation cycle.

To assess the frequency of collisions capable of producing a crater that can penetrate the outermost comet layer and expose ices, we consider a depth of $h=10$ m. The relationship between depth $h$ and diameter $D$ of a crater for asteroids has a mean value of $h/D = 0.15$ \citep{Marchi2015}. Thus, we need to consider craters with $D = 66$ m. Using Eq. \ref{eq:Crater}, we determine that the impactor producing such a crater on 1-15 km targets has a size of approximately 4 m. We use Eq. \ref{eq:frecuencias}, starting the summation from 4 m, and plot the timescale in the bottom panel of Fig. \ref{fig:FrecQH}. We find that the timescale increases significantly at the small end due to the limited range of impactor sizes considered. However, for larger-sized bodies, the set of possible impactor sizes expands, resulting in a slight reduction in the timescale. The shortest timescale is associated with the S32 scenario, reaching down to $2\times10^6$ years, but it still exceeds the dynamic lifetimes of quasi-Hildas in the region.

\section{Conclusions and discussions}

In this study, we have performed a collisional study of the Hilda asteroids and their potential for activating quasi-Hildas through impacts. Here, we summarize our key findings:

\begin{itemize}
    \item We obtained the SFDs of the observed Hilda and quasi-Hilda populations. The SFD of the Hilda population exhibits a cumulative slope of $-2.03$, similar to Jupiter's Trojan asteroids. In contrast, the quasi-Hilda SFD displays a two-slope power law, with a slope of $-2.4$ for bodies between 3 km and 6 km, and $-1.21$ for bodies between 6 km and 20 km.
    \item We explored the collisional evolution of Hilda asteroids over a 4 Gyr period. We found that scenarios assuming minimal catastrophic collisional activity in the multikilometer-sized Hilda asteroids produced the best fits to the observed SFDs. The slopes of the initial SFDs for subkilometer-sized bodies were found to steepen over time due to the fragmentation of a small number of multikilometric sized bodies. 
    \item We analyzed the impactor and cratering events on the largest member of the Hilda asteroids, 334 Chicago. Our results show that this asteroid experienced minimal impacts from large bodies over the integration period. The largest crater produced on this asteroid has between 20 and 32 km in size. Moreover, the crater SFD reflects the slope of the initial SFD in the subkilometer range. Therefore, a mission aimed at observing this asteroid and in particular the craters on its surface would be of utmost importance to restrict the original population of the Hildas and their origin.
    \item Impacts on quasi-Hildas can lead to observable activity, with even modest collisions capable of doubling the brightness of these objects. However, the timescale of impact-triggered activity is notably larger than the dynamical lifetime of quasi-Hildas. Despite their potential to induce activity and the stochastic nature of such events, the collisions between Hildas and quasi-Hildas are unlikely to be a primary mechanism for the observed phenomena.     
\end{itemize}

The results of this study reaffirm the findings of previous works in the literature regarding the limited overall evolution experienced by the Hilda asteroids. However, it is important to note that this analysis was conducted considering only collisions between Hilda asteroids. Other small-body populations, such as the Jupiter Trojans or MB asteroids, may also interact collisionally with Hilda and quasi-Hilda objects, although the intrinsic collisional probability with these populations is an order of magnitude lower than that between Hilda objects \citep{dahlgren1998study,dell2001updated}. On one hand, the contribution of the Jupiter Trojan population may be neglected. We can estimate a timescale of impacts between pairs of Hilda targets with size $D_\text{H}$ and Trojan projectiles $D_\text{T}$ as the inverse of $\frac{P_\text{i}}{4}(D_\text{T}+D_\text{H})^2N(>D_\text{T})$, with $P_\text{i}=0.27\times 10^{-18}$ yr$^{-1}$ km$^{-2}$ \citep{dell2001updated}.  $N(>D_\text{T})$ is the SFD of the Trojan population \citep{di2019dynamical}, which is assumed to be primarily primordial in the multikilometer size range \citep{marschall2022implications}. Based on this, the catastrophic collisions with Trojan projectiles capable of substantially modifying the Hilda SFD occur on timescales comparable to the age of the Solar System, while small cratering impacts occur on times greater than $10^7$ years. On the other hand, the contribution from the MB could be assessed within the context of a study focused on the collisional evolution of the MB itself, which is beyond the scope of this work. This is a non-trivial matter, as the MB is a large and complex structure with distinct physical and dynamical properties, and can be divided into at least three regions: the inner, middle, and outer belts \citep{deElia2007,Cibulkova2014,Zain2020}. This is particularly relevant for the Hilda population, as its members are more likely to collide with bodies from the outer belt, given that the mean perihelion distance of Hilda objects is approximately 3.13 au. Specifically, a collisional evolution study focused on the outer belt could provide the currently unconstrained SFD in the sub-kilometer size range, which is necessary for calculating the impact timescales between outer belt asteroids and quasi-Hildas. This represents an interesting area for future research on the collisional evolution of small solar system bodies.

\begin{acknowledgements}
This research was performed using computational resources of Facultad de Ciencias Astronómicas y Geofísicas de La Plata (FCAGLP) and Instituto de Astrofísica de La Plata (IALP), and received partial funding from Universidad Nacional de La Plata (UNLP) under PID G172. RGH gratefully acknowledges financial support by CONICET through PIP 112-202001-01227 and San Juan National University by a CICITCA grant for the period 2023-2024.
\end{acknowledgements}

\bibliographystyle{aa} 
\bibliography{biblio} 

\end{document}